\documentclass[twocolumn,showpacs,amsmath,amssymb]{revtex4-1}
\usepackage{graphicx}
\begin{document}
\title{\Large \bf Type N Spacetimes as Solutions of Extended  New Massive Gravity}
\author{\large Haji Ahmedov}
\address{T\"{U}B\.{I}TAK-B\.{I}LGEM, Research Institute for Fundamental Sciences, 41470  Gebze, Turkey}
\author{\large Alikram N. Aliev}
\address{ Do\v{g}u\c{s} University, Department of Mathematics, Kad{\i}k\"{o}y, 34722  Istanbul, Turkey}
\date{\today}

\begin{abstract}

We study algebraic type N spacetimes in the extended new massive gravity (NMG), considering  both the Born-Infeld model (BI-NMG) and the model of NMG with any finite order curvature corrections.  We show that  for these spacetimes, the field equations of BI-NMG take the form of  the massive (tensorial)  Klein-Gordon type equation, just as it happens for ordinary  NMG. This fact enables us to obtain the type N solution to  BI-NMG, utilizing the general type N solution of NMG, earlier found in our work. We also obtain type N solutions to  NMG  with all finite order curvature corrections  and show that, in contrast to BI-NMG, this model admits the critical point solutions, which are   counterparts of ``logarithmic"  AdS pp-waves   solutions of NMG.

\end{abstract}


\maketitle

\section{Introduction}

In striving to search for a toy model of quantum gravity, three-dimensional gravity theories have been  an active area of investigations for many years. In a recent development, a new intriguing theory of three-dimensional massive gravity was proposed in \cite{bht1} where a particular higher-derivative term, added to the  Einstein-Hilbert action, confers dynamical degree of freedoms on the theory. This theory is  referred to as {\it new massive gravity} (NMG) and, in contrast to the earlier known theory of  topologically massive gravity (TMG) \cite{djt, deser}, it is a  parity-preserving theory whose propagating  degrees of freedom involve two massive spin-2 modes. The linearized equations of NMG in Minkowski vacuum turn out to be equivalent to those of the unitary Fierz-Pauli theory  for a free massive spin-2 field \cite{bht1}. Meanwhile, unitarity of linearized NMG in AdS vacuum implies a negative central charge in the dual conformal field theory (CFT) on the boundary \cite{bht2}. Detailed  studies of the unitarity properties of NMG as well as the discussion of various attempts to reconcile the unitarity conflict in the bulk/boundary theories can be found in \cite{deser1, tekin1, gregory, liu1, daniel1, tekin2}.

The appearance of NMG almost immediately  raised an important  question of whether one can obtain a sensible theory of three-dimensional gravity by adding the next order curvature invariants to the action of NMG. The first step towards this direction was made in \cite{sinha1}, where the theory of NMG was  reproduced by assuming  the existence of the dual CFT  and a simple holographic c-theorem. Within this approach, NMG  was extended to involve cubic and quartic curvature  corrections. The further use of this approach resulted in the extension of NMG involving  an arbitrary (finite) number of curvature corrections \cite{pau}. An interesting Born-Infeld (BI) type gravity theory in three dimensions was proposed in \cite{tekin3}. In a certain sense, this theory  is an extension of NMG to include infinite order curvature corrections. In the weak curvature limit, it  matches  NMG  as well as its extension  that involves cubic curvature  corrections, obtained earlier within the holography approach \cite{sinha1}. This in turn indicates that BI-NMG must  allow  simple c-functions. In \cite{tekin4}, it was shown that this is indeed the case with the assumption  of the null-energy condition. However, the conflict between the bulk unitarity and the unitarity of the boundary CFT, existing in NMG and in the extended NMG with   cubic and quartic curvature  corrections, retains in BI-NMG as well, albeit it becomes less severe \cite{tekin5}. It is also interesting to note that the theory of  BI-NMG  naturally arises as a boundary counterterm in the anti-de Sitter space (AdS$_4$)  within the AdS/CFT correspondence \cite{sinha2}. As for the exact solutions to the theory, it was  shown that while BI-NMG admits the usual AdS pp-wave solution just like NMG, it does not admit  a logarithmic type solution at the critical value of the parameters, in contrast to NMG \cite{mohsen}. Various AdS and warped AdS types  black hole solutions to the extended NMG have been discussed in \cite{nam, ahmad}.

Recently, a novel approach to NMG, built up on the use of a first-order differential operator appearing in TMG, was developed in  \cite{ah1}. With this differential operator, the field equations of TMG  can be interpreted as a massive (tensorial) Dirac type equation.  Assuming that the first-order differential (TMG) operator somewhat  lies at the ``heart"  of the NMG theory as well, it was shown  that the field equations of NMG reduce to the form of a massive (tensorial) Klein-Gordon type equation, involving the square of  this operator. It turns out that  for algebraic types D and N spacetimes  with constant scalar curvature, the relationship between the  Klein-Gordon type
form of the NMG equations and the TMG  equations closely resembles the relationship between the usual Klein-Gordon and Dirac equations.  That is, in the case under consideration,  TMG can be considered as the ``square root" of NMG. This in turn means that any type D or N  solution of TMG  is  also a solution to  NMG (after adjusting the associated parameters). Altogether, this approach greatly facilitates the study of all types D and N  solutions of  NMG \cite{ah2, ah3}. For other works on exact solutions of NMG see, for instance, \cite{troncoso, giri, mg, nam1}.

The purpose of this Letter is to employ the novel approach of \cite{ah1} to exploring type N spacetimes in the  extended NMG. In Sec.II we briefly discuss the structure of the BI-NMG model and present the  equations of motion. In Sec.III we describe the dynamics of type N spacetimes  in  BI-NMG, using the language of a ``semi-null"  triad of real vectors. Next, we show that the use of the first-order differential operator, emerging in TMG,  renders the field equations of BI-NMG to take the form of  the massive (tensorial)  Klein-Gordon type equation  (accompanied by an associated constraint equation).  In Sec.IV we obtain the type N solution  to  BI-NMG  by using the general type N solution of NMG, earlier found in \cite{ah3}. In Sec.V we discuss the equations of motion for type N spacetimes in NMG with all finite order curvature corrections. Here we show that this model admits the counterparts of  the critical point solutions of NMG, in contrast to the BI-NMG model.

\section{Basics of BI-NMG}

As we have mentioned above, the extension of NMG involving infinite order curvature corrections was proposed in \cite{tekin3}.  This  model is described by the Born-Infeld type action
\begin{eqnarray}
S&=&\frac{1}{16\pi G}  \int d^3 x \sqrt{-g}\,\left[4 m^2 \left(\sqrt{\det A} -1\right) -2\Lambda\right],
\label{biaction}
\end{eqnarray}
where  $ m $ is the mass parameter, $\Lambda $  is the cosmological constant,
$ A $  denotes the tensor
\begin{eqnarray}
A_\mu^{\,\nu} &=& \delta_\mu^{\,\nu} - \frac{1}{m^2} \, G_\mu^{\,\nu}\,
\label{atensor}
\end{eqnarray}
and the Einstein tensor $ G_{\mu\nu} = R_{\mu\nu}  - \frac{1}{2}\,g_{\mu\nu} R  $. Expanding this action in the weak curvature regime, or equivalently in powers of the parameter  $  1/m^2 $  up to  order $  (1/m^2)^{N-1}, \,$  we can obtain the  NMG action with all finite order curvature corrections (see also \cite{tekin3, pau}). In particular, at the quadratic order $ (N=2) $ we arrive at the usual  NMG action
\begin{eqnarray}
S &=& \frac{1}{16\pi G} \int d^3x \sqrt{-g}\left(R - 2 \Lambda -\frac{1}{m^2}\, K \right),
\label{nmgaction}
\end{eqnarray}
where the scalar  $ K $ is given by
\begin{eqnarray}
K &=& R_{\mu\nu}R^{\mu\nu}- \frac{3}{8} \,R^2\,.
\label{k}
\end{eqnarray}
The equations of motion for BI-NMG are obtained by varying  action (\ref{biaction}) with respect to the spacetime metric \cite{tekin4, nam}. As a consequence, we have
\begin{eqnarray}
B_{\mu\nu; \lambda}^{~~~~\ ;\lambda}+ B_{; \mu; \nu} - 2 \,B_{\lambda(\mu; \nu)}^{~~~~~~ ;\lambda} + 2 \,B^{~\lambda}_{(\mu}  R_{\nu )\lambda} - B
R_{\mu\nu}
\nonumber \\[2mm]
+ g_{\mu\nu}\left[B^{\mu\nu}_{\ \ \ ;\mu;\nu} -  B_{;
\lambda}^{ \ \ ;\lambda}
+  2m^2\left(\sqrt{\det{A}} - 1\right) -\Lambda \right]
  &=& 0\,,
\nonumber\\
\label{feq}
\end{eqnarray}
where the semicolon denotes  covariant differentiation and we have used the notations
\begin{equation}
B_{\mu\nu}= \sqrt{\det A } \ \left(A^{-1}\right)_{\mu\nu}, \ \ \ \ \ B= B_\mu^{\ \mu}\,.
\label{notations}
\end{equation}
The trace of  equation (\ref{feq}) is given by
\begin{eqnarray}
B_{\mu\nu}^{~~ ;\mu;\nu} - B_{;\mu}^{~ ;\mu} +
2 B_{\mu\nu} R^{\mu\nu} -  B R
\nonumber \\
+ 3\left[ 2m^2\left(\sqrt{\det{A}}-1\right) -\Lambda
\right] &=& 0\,.
\label{trfeq}
\end{eqnarray}
We note that the equations of motion for NMG with finite order curvature corrections can be  obtained  by  successive  expansion of equations (\ref{feq}) and  (\ref{trfeq}) up to  order $  (1/m^2)^{N-1} $.

\section{Dynamics of Type N Spacetimes in BI-NMG}

It is convenient to begin by  introducing a semi-null  triad of real vectors  $ \{ l_{\mu}\,,  n_{\mu}\,, m_{\mu}\} $, where  $  l_{\mu}$  and   $ n_{\mu} $ are null vectors and $ m_{\mu} $ is a spacelike vector. They  satisfy the relations
\begin{eqnarray}
l_{\mu}  n^{\mu} &= & 1\,,~~~~~ m_{\mu}  m^{\mu}=1\,,
\label{norms}
\end{eqnarray}
whereas all the remaining  contractions vanish identically. Clearly, the spacetime metric can be written in terms of these vectors as follows
\begin{eqnarray}
g_{\mu\nu} &= & 2 l_{(\mu} n_{\nu)} + m_{\mu} m_{\nu}\,.
\label{basemetric}
\end{eqnarray}
It is also known that the use of  these vectors in the Petrov-Segre classification of three-dimensional spacetimes \cite{chow} renders the  traceless Ricci tensor
\begin{eqnarray}
S_{\mu\nu} &=& R_{\mu\nu}-\frac{1}{3}\,g_{\mu\nu} R
\label{trlessricci}
\end{eqnarray}
of type N spacetimes to take the simple canonical form
\begin{eqnarray}
S_{\mu\nu}= l_{\mu}l_{\nu}\,.
\label{Nricci1}
\end{eqnarray}
With this in mind,  equation (\ref{atensor}) can be written in the form
\begin{eqnarray}
A_{\mu\nu} & = & q\, g_{\mu\nu}-  \frac{1}{m^2} \,S_{\mu\nu}\,,
\label{atensor1}
\end{eqnarray}
where
\begin{equation}
q=1+\frac{R}{6 m^2}\,\,.
\label{q}
\end{equation}
Using  metric (\ref{basemetric}) and taking into account the property of the basis vectors in (\ref{norms}), it is easy to find the representation of $ A_{\mu}^{~\nu} $ in the  chosen basis. It is given by
\begin{eqnarray}
A_\mu^{\ \nu}\, l_\nu &=& q     l_\mu\,, \nonumber\\
A_\mu^{\ \nu} \,n_\nu &=& q n_\mu -\frac{1}{m^2} \,l_\mu\,, \nonumber\\
A_\mu^{\ \nu}\,s_\nu &= & q s_\mu\,.
\label{rep}
\end{eqnarray}
This in turn  results in the matrix
\begin{eqnarray}
A &=&\left(
  \begin{array}{ccc}
    q ~ & 0~ & ~0 \\[2mm]
    -1/m^2~ & q~ &~ 0 \\[2mm]
     0 ~& 0~  & q~ \\[2mm]
\end{array}
\right)
\label{matrix}
\end{eqnarray}
with
\begin{eqnarray}
\det A &=& q^3\,.
\label{det}
\end{eqnarray}
Comparison  of equations (\ref{biaction}) and (\ref{det}) reveals the inequality
\begin{equation}
1+\frac{R}{6 m^2} > 0 \,.
\label{qq}
\end{equation}
Using now equations (\ref{norms}) and (\ref{basemetric}) one can also invert (\ref{atensor1}) to obtain a matrix whose components are given by
\begin{eqnarray}
(A^{-1})_{\mu\nu} &=& \frac{1}{q}\left( g_{\mu\nu}+
\frac{1}{q\, m^2} \,S_{\mu\nu}\right).
\label{inverse}
\end{eqnarray}
Substitution of this expression, along with (\ref{det}), into equation (\ref{notations}) yields
\begin{eqnarray}
B_{\mu\nu} &=& \sqrt{q}\, g_{\mu\nu} + \frac{1}{\sqrt{q}\,
 m^2} \,S_{\mu\nu}\,,~~~ B= 3 \sqrt{q} \,\,.
 \label{bb}
\end{eqnarray}
It is straightforward to show that these quantities fulfil the differential constraint given by
\begin{eqnarray}
B_{\mu\nu}^{\ \ \ ;\nu} &= & B_{;\mu}\,.
 \label{difconst}
\end{eqnarray}
Indeed, with equations (\ref{q}) and  (\ref{bb}) we find that
\begin{eqnarray}
 B_{\mu\nu}^{\ \ \ ;\nu}- B_{;\mu} &=& - \frac{l_{\mu}}{12 m^4 q^{3/2}}\,\, l^\nu R_{;\nu} \,,
 \label{diff}
\end{eqnarray}
where we have used the contracted Bianchi identity
\begin{equation}
{S_{\mu\nu}}^{ ;\nu}=\frac{1}{6}\, R_{;\mu}\,
\label{contrbian}
\end{equation}
and equation (\ref{Nricci1}).  Contracting now both sides of   (\ref{contrbian}) with $ l^{\mu} $,  we obtain $ l^\nu R_{;\nu}=0 $. That is, the right-hand side of (\ref{diff}) vanishes, leaving us with (\ref{difconst}).

It is important to note that the differential  constraint in (\ref{difconst})
greatly simplifies the further  description of type N spacetimes in  BI-NMG. With this constraint, the trace equation (\ref{trfeq})  takes the most simple form
\begin{eqnarray}
\sqrt{q} &=& 1+\frac{\Lambda}{2 m^2}\,.
\label{constr1}
\end{eqnarray}
This in turn, by means of (\ref{q}), yields
\begin{eqnarray}
R & =&  6\Lambda \left(1+\frac{\Lambda}{4m^2}\right)\,,
 \label{trconst}
\end{eqnarray}
thereby relating the scalar curvature of type N spacetimes  to the cosmological and  mass parameters of the BI-NMG model.

Next, following our previous \cite{ah1} work, we introduce  a first-order differential operator $ {D\hskip -.25truecm \slash} \,$ defined as
\begin{eqnarray}
{D\hskip -.25truecm \slash}\,B_{\mu\nu} &= & \frac{1}{2}\left( {\epsilon_{\mu}}^{\alpha\beta} B_{\nu\alpha ;\beta} + {\epsilon_{\nu}}^{\alpha\beta} B_{\mu\alpha ;\beta} \right),
\label{doper1}
\end{eqnarray}
where $ B_{\mu\nu} $ is  an arbitrary symmetric tensor (not necessarily that given in  equation (\ref{bb})) and the Levi-Civita tensor $\epsilon_{\mu\alpha\beta}= \sqrt{-g} \,\varepsilon_{\mu\alpha\beta} $, where we  take  $\varepsilon_{012}=1 $. It is not difficult  to show that if the condition
\begin{eqnarray}
{\epsilon_{\lambda}}^{\mu\nu} {\epsilon_{\mu}}^{\alpha\beta} B_{\nu\alpha ;\beta} &= &  0
\label{difcond2}
\end{eqnarray}
holds,  one can introduce the ``truncated"  definition for this operator. It is given by
\begin{eqnarray}
{D\hskip -.25truecm \slash\,} B_{\mu\nu} &= & {\epsilon_{\mu}}^{\alpha\beta} B_{\nu\alpha ;\beta} \,.
\label{doper2}
\end{eqnarray}
It is also straightforward to show that condition (\ref{difcond2}) is equivalent to that given in (\ref{difconst}). That is,  for the $ B_{\mu\nu}$  tensor (\ref{bb}) it is legitimate to use  the  definition in (\ref{doper2}).

It is remarkable that the truncated definition  also works for the secondary action of $ {D\hskip -.25truecm \slash\,} $ on the  tensor  $ B_{\mu\nu} $  in (\ref{bb}).  Indeed, using first equation (\ref{bb})  we see that
 \begin{eqnarray}
 {D\hskip -.25truecm \slash\,} B_{\mu\nu}   &=&  \frac{1}{\sqrt{q}\,
 m^2} \,\,{D\hskip -.25truecm \slash\,}S_{\mu\nu}\,.
 \label{dbb}
\end{eqnarray}
Then, we recall that the  tensor $ {D\hskip -.25truecm \slash\,}S_{\mu\nu} $, as shown in \cite{ah1}, can be expressed in terms of the Cotton tensor  $ C_{\mu\nu} $ as follows
\begin{eqnarray}
{D\hskip -.25truecm \slash\, } S_{\mu\nu} &= & - C_{\mu\nu}\,,
 \label{cot1}
\end{eqnarray}
where
\begin{eqnarray}
C_{\mu\nu} &= & {\epsilon_{\mu}}^{\alpha\beta}\left(R_{\nu\beta} - \frac{1}{4}\, g_{\nu\beta} R\right)_{;\alpha}\,
\label{cotton}
\end{eqnarray}
Since the  Cotton tensor fulfils the condition (\ref{difconst}), comparing equations (\ref{dbb}) and (\ref{cot1}),  we see that  the same is true for the tensor $ {D\hskip -.25truecm \slash\,} B_{\mu\nu} $ as well. Thus, as stated above, we have
\begin{eqnarray}
{D\hskip -.25truecm \slash\,}^2  B_{\mu\nu} &=& {\epsilon_{\mu}}^{\alpha\beta} \left({D\hskip -.25truecm \slash\,} B_{\nu\alpha}\right)_{;\beta}\,.
\label{secdd}
\end{eqnarray}
Inserting  now equation (\ref{doper2}) into the right-hand side of this expression and performing some calculations, we find that
\begin{eqnarray}
{D\hskip -.25truecm \slash\,}^2  B_{\mu\nu} &=& {B_{\mu\nu ;  \lambda}}^{;\lambda} - {B_{\lambda(\mu ;\nu)}}^{;\lambda}\,.
\label{nmgdirac2}
\end{eqnarray}
We recall that by definition of the Riemann tensor and with the use of equation (\ref{difconst}),  we have the relation
\begin{eqnarray}
{B_{\lambda \mu; \nu}}^{;\lambda}= {R_{\lambda\nu\mu}}^{
\tau} B_\tau^{\ \lambda}+R_\mu^{\ \ \lambda} B_{\lambda\nu} + B_{; \mu;\nu}\,.
\label{riemann1}
\end{eqnarray}
Using this relation in (\ref{nmgdirac2}) and substituting the result into  equation (\ref{feq}), with equation (\ref{difconst}) in mind, we transform it into the form that involves the square of the operator $ {D\hskip -.25truecm \slash\,} $. After some rearrangements, we have
\begin{eqnarray}
{D\hskip -.25truecm \slash\,}^2  B_{\mu\nu}- {R_{\lambda\nu\mu}}^{
\tau} B_\tau^{\ \lambda} +  R_{\lambda(\mu} B^\lambda_{\nu)} - B R_{\mu\nu}
\nonumber \\[2mm]
+ g_{\mu\nu}\left[ 2m^2\left(\sqrt{\det{A}} - 1\right) -\Lambda \right]
  &=& 0\,.
\label{feq1}
\end{eqnarray}
Next, with the use of the expression for the Riemann tensor in three dimensions
\begin{equation}
R_{\mu \nu \alpha \beta}  =  2\left(R_{\mu [\alpha}g_{\beta]\nu} - R_{\nu [\alpha} g_{\beta]\mu}
-\frac{R}{2}\, g_{\mu[\alpha}g_{\beta]\nu}\right),
\label{rimtoric}
\end{equation}
where the square brackets denote antisymmetrization, equation (\ref{feq1}) takes the form
\begin{eqnarray}
{D\hskip -.25truecm \slash\,}^2  B_{\mu\nu} -  R_{\lambda(\mu} B^\lambda_{\nu)}
+\frac{1}{2}\,R \left(B_{\mu\nu} - g_{\mu\nu} B \right)
\nonumber \\[2mm]
+ g_{\mu\nu}\left[2 m^2(\sqrt{\det{A}}-1) - \Lambda
+ R_{\mu\nu} B^{\mu\nu} \right] &=& 0\,.
\label{feq2}
\end{eqnarray}
Finally, substituting into this equation  the explicit form of the tensor $ B_{\mu\nu} $ given in (\ref{bb}) and taking into account (\ref{Nricci1}) and (\ref{det}), we arrive at the massive (tensorial) Klein-Gordon type equation
\begin{eqnarray}
{D\hskip -.25truecm \slash\,}^2  S_{\mu\nu} = m^2 S_{\mu\nu}\,,
\label{kg1}
\end{eqnarray}
which, augmented by the constraint equation (\ref{trconst}), describes the full dynamics of type N spacetimes in BI-NMG.  Thus, {\it we again have the massive  Klein-Gordon type equation, just as for ordinary new massive gravity} \cite{ah1}.  This means that one can easily obtain type N solutions to BI-NMG  using those of NMG \cite{ah3}. Below, we discuss these solutions.

\section{Solutions to BI-NMG}

In our previous work \cite{ah3}, we have found that the type N  metric
\begin{equation}
ds^2  =   d\rho^2 + \frac{2 du dv}{\nu^2-\beta^2}+\left[Z(u,\rho) - \frac{v^2}{\nu^2-\beta^2} \right]du^2,
\label{unimetric}
\end{equation}
with the metric functions
\begin{eqnarray}
Z(u,\rho) & = & \frac{1}{\sqrt{\nu^2-\beta^2}}\left[\cosh(\mu\rho)\, F_1(u)+ \sinh(\mu\rho)\, F_2(u)\right. \nonumber \\[2mm]  & & \left.
+ \cosh(\nu\rho)\,  f_1(u) + \sinh(\nu\rho)\,  f_2(u)\right],
\label{zetunimt1}
\end{eqnarray}
$\beta = \nu \tanh(\nu\rho)$ or   $\beta = \nu \coth(\nu\rho)\,$  is the general solution to the field equations of NMG
\begin{eqnarray}
{D\hskip -.25truecm \slash\,}{^2}S_{\mu\nu} & = & \mu^2 S_{\mu\nu}\,, ~~~~~~
\mu^2  =   {\tilde m}^2 +\frac{\nu^2}{2}\,,
\label{Nkg}
\end{eqnarray}
accompanied by the constraint equation
\begin{eqnarray}
\lambda & = & -\nu^2 \left(1+\frac{\nu^2}{4{\tilde m^2}}\right),
\label{cosparam}
\end{eqnarray}
where $ \lambda $ is the cosmological parameter  and  ${\tilde m}$ is the mass
 parameter of NMG. Here we have also used the notation  $ \nu^2 = - R/6 \,$. As it was shown in  \cite{ah3},  solution (\ref{unimetric}) is indeed characterized by three arbitrary functions. One of  two extra functions $ f_1(u) $ and  $ f_2(u) $, appearing in the metric, is  redundant and it can be  removed away by  appropriate coordinate  transformations. However, for some purposes it is convenient to keep in the metric both of these functions. Moreover, this solution  does not possess the null Killing isometry, but this isometry emerges in the limiting case $ \beta^2 =  \nu^2  $, where the solution describes  AdS pp-waves \cite{ah3, giri}.

In \cite{ah3}, we considered two other critical point solutions as well: Namely, the solution corresponding to the case $ \mu^2 =\nu^2 $ ($ {\tilde m}^2=\nu^2/2 $), for which  the metric function in (\ref{unimetric}) is given by
\begin{eqnarray}
Z(u,\rho) & = & \frac{1}{\sqrt{\nu^2-\beta^2}} \left \{\cosh(\mu\rho) \left[\rho F_1(u)+f_1(u)\right] \right. \nonumber \\[2mm]  & & \left.
+ \sinh(\mu\rho) \left[\rho F_2(u)
+f_2(u)\right]\right\},
\label{zetcritical}
\end{eqnarray}
instead of (\ref{zetunimt1}), and the solution in the case $ \mu^2 =0 $  ($ {\tilde m^2} =-\nu^2/2 $),  for  which we have
\begin{eqnarray}
Z(u,\rho) & = & \frac{1}{\sqrt{\nu^2-\beta^2}}\left[F_1(u)+ \rho\, F_2(u)\right. \nonumber \\[2mm]  & & \left.
+ \cosh(\nu\rho)\,  f_1(u) + \sinh(\nu\rho)\,  f_2(u)\right].
\label{zetmu0}
\end{eqnarray}
The wave profile analysis of these solutions  reveals the appearance of  additional logarithmic modes, which becomes evident when making the  coordinate transformation $\rho \rightarrow \nu^{-1} \ln y $. That is, the  solutions under consideration are of type N counterparts of ``logarithmic"  AdS pp-waves   solutions, earlier considered in \cite{giri}.

Comparing now the field equations in (\ref{kg1}) and  (\ref{Nkg}),  with equation (\ref{trconst}) in mind, it is not difficult to see that for
\begin{eqnarray}
\mu^2 & =&  m^2 \,,~~~~~~\nu^2  =  - \Lambda \left(1+\frac{\Lambda}{4m^2}\right)\,,
\label{trconsttt}
\end{eqnarray}
the general type N solution of NMG  is also subject to  BI-NMG. This is confirmed by computer calculations as well.

As for the  counterparts of solutions with  (\ref{zetcritical}) and (\ref{zetmu0}) we note the following: (i)  At the critical point  $ m^2 = \nu^2= -R/6\, $, the determinant in (\ref{det}) vanishes. That is,  at this point the theory fails to exist itself. (ii)  The critical value $ \mu^2 =0 $ in equation (\ref{Nkg}) implies the vanishing of the mass parameter in (\ref{kg1}), $ m^2=0 $. However, this is not allowed in BI-NMG. We conclude that {\it the counterparts of the solutions given in (\ref{zetcritical}) and  (\ref{zetmu0}) do not exist in  the Born-Infeld   extension of NMG}.  Similar result was earlier obtained in \cite{mohsen} for the AdS pp-wave solutions with logarithmic modes.

\section{Solutions to NMG  with  finite order curvature corrections}

The equations of motion for type N spacetimes in NMG with all finite order curvature corrections can be obtained from those given above, by employing an appropriate expansion procedure. In the weak curvature limit, the constraint equation (\ref{constr1})  is also appropriate for NMG  with $N$-order curvature corrections, provided that on  the left-hand side of (\ref{constr1}) we perform the Taylor expansion. Thus,  we have
\begin{eqnarray}
\left(\sqrt{q}\right)_N &=& 1+\frac{\Lambda}{2 m^2}\,\,,
\label{constr2}
\end{eqnarray}
where
\begin{eqnarray}
\left(\sqrt{q}\right)_N & = & \sum_{n=0}^{N}\, \frac{1}{n!}\,\left(
    \frac{d^n}{dx^n}\sqrt{q}\right)_{x=0}\, x^n\,.
\label{taylor}
\end{eqnarray}
We recall that  $ q=1+ x $ and $ x=R/6m^2 $. The expansion  is taken up to  order $  (1/m^2)^{N} $\,. For  $ N=2 $, from (\ref{constr2}) we find the relation
\begin{eqnarray}
\Lambda & = &\frac{R}{6}\left(1-\frac{R}{24 m^2}\right),
\label{nmgconstr}
\end{eqnarray}
which is precisely the same  as  that  obtained in \cite{ah3}  for type N spacetimes  in  NMG.

Similarly, one can present the Klein-Gordon type equation  for NMG  with  $N$-order curvature  corrections.  Indeed,  using  equation (\ref{bb}) in  (\ref{feq2}) and performing some calculations,  with equations (\ref{Nricci1}) and (\ref{det}) in mind,  we obtain that
\begin{eqnarray}
{D\hskip -.25truecm \slash\,}^2  S_{\mu\nu} & = & \mu_N ^2 S_{\mu\nu}\,,
\label{rnkg}
\end{eqnarray}
where  the mass parameter is given by
\begin{eqnarray}
\mu_N ^2 = m^2 \,\frac{\left(1/\sqrt{q}\right)_{N-1}}{\left(1/\sqrt{q} \right)_{N-2}}
\label{mun}
\end{eqnarray}
and the subscripts $ N-1 $ and $ N-2 $ at  the round  brackets  imply the corresponding Taylor expansions (see Eq.(\ref{taylor})).

Now, it is not difficult to see that the general type N solution  of equation (\ref{Nkg}) (see  metric (\ref{unimetric})) is also  the solution  to equation (\ref{rnkg}),  provided that $ \mu^2= \mu_N ^ 2 $ and   the parameter $ \nu^2 = - R/6 \,$ is determined by equation (\ref{constr2}). Moreover, it turns out that the NMG  model with $N$-order curvature corrections, unlike the BI-NMG  model,  admits the counterparts of  the critical point solutions of NMG, given above  with (\ref{zetcritical}) and  (\ref{zetmu0}). Let us now discuss these solutions:

(i) Clearly, the critical solution  with  (\ref{zetcritical}) exists when the condition  $ \mu_N ^ 2= \nu^2 $  is fulfilled. This  condition can also be written in the form
\begin{eqnarray}
\left(1/\sqrt{q}\right)_{N-1} + x\,\left(1/\sqrt{q}\right)_{N-2} &=&0\,.
\label{creq1}
\end{eqnarray}
Next, we solve this  algebraic equation separately for each even and odd values of $N $.  We find that the for each even value of $ N $  there exists only one real root. For instance, we have
\begin{eqnarray}
N &=& 2\,,~~~~~~~~ x= -2\,,\nonumber \\
N &=& 4\,,~~~~~~~~ x \simeq  -1.30\,,\nonumber \\
N &=& 6\,,~~~~~~~~ x \simeq  -1.17\,,
\label{sols1}
\end{eqnarray}
where the first line corresponds to the solution for ordinary  NMG (see Eq.(\ref{zetcritical})). We see that as  $ N $ increases  the value of $ x $  grows as well. The further numerical analysis of (\ref{creq1}) shows that  $ x\rightarrow -1$ for large enough even values of   $ N $.

Meanwhile, for each odd value of $ N $,  we find  two different real roots. That is, we have
\begin{eqnarray}
N &=& 3\,,~~~~~~~~x \simeq  5.46\,,~~~~~~ x \simeq  -1.46 \,,\nonumber \\
N &=& 5\,,~~~~~~~~x \simeq  2.96\,,~~~~~~   x \simeq  -1.22 \,,\nonumber \\
N &=& 7\,,~~~~~~~~x \simeq  2.28\,,~~~~~~ x \simeq  -1.15 \,.
\label{sols2}
\end{eqnarray}
Again, we see that with the growth of $ N $, the value of the positive  root  decreases, whereas the value of the negative  root  increases. The detailed numerical analysis reveals that for sufficiently large odd values of $ N $, we have  $ x\rightarrow 1 $  and  $ x\rightarrow -1 $.

(ii) The critical solution with metric function (\ref{zetmu0})  corresponds to the case $ \mu_N ^ 2= 0 $. Equivalently, we have the condition
\begin{eqnarray}
\left(1/\sqrt{q}\right)_{N} &=&0\,.
\label{creq2}
\end{eqnarray}
Solving this equation numerically at each order of the expansion,  we find that it has no real roots for odd values $ N $. On the other hand, for even values of $ N $ we have the following roots
\begin{eqnarray}
N &=& 2\,,~~~~~~~~ x= 2\,,\nonumber \\
N &=& 4\,,~~~~~~~~ x \simeq  1.53\,,\nonumber \\
N &=& 6\,,~~~~~~~~ x \simeq  1.37\,.
\label{sols1}
\end{eqnarray}
We note that the root $ x= 2 $ matches the critical solution  for NMG (see Eq.(\ref{zetmu0})). Continuing the numerical analysis, we see that  $ x\rightarrow 1 $ for large enough even values of $ N $.

It is worth noting once again that the model of BI-NMG works for any values of the parameter $ x= R/6m^2 $ that fulfills  condition (\ref{qq}). For $ x= -1 $, the model fails to exist as the  determinant (\ref{det}) vanishes at this point. Our analysis shows that the scalar function $ q= 1+x $ plays a defining role in the theory. For $ |x|< 1 $,  performing  the Taylor expansion of the function $ 1/\sqrt{q} $,  we arrive at the extension of  NMG  with (finite) $N$-order curvature corrections. It is amazing  that  the finite series  expansion turns out to be the well-defined extension of NMG for arbitrary values of  $ x $ as well, admitting the critical point solutions, which are absent in BI-NMG.

\section{Conclusion}

The use of the first-order  TMG operator in the structure of NMG reveals an intimate relation between the two theories, reminiscent of the relation between the ordinary  Klein-Gordon and Dirac equations \cite{ah1}. This fact played a defining role in finding the  full set of types  D and N exact solutions to NMG \cite{ah2,ah3}. In this Letter, we  applied the same line of thought   to the extended models of NMG, namely to its Born-Infeld extension  and to the  extension involving any finite order curvature corrections.  Though  at first sight, the field equations of BI-NMG  look very  austere, we have shown that for type N spacetimes they reduce to a  simple  Klein-Gordon type form,  with the use of a semi-null  triad of real vectors  and  the first-order TMG operator. That is, we again have obtained the massive Klein-Gordon type equation involving the square of the TMG operator, just as in the case of NMG.  Adjusting the parameters of this equation and those of the equation appearing in NMG, we  managed to obtain type N solution of BI-NMG by using the general type N solution of NMG. We have also shown that the critical point solutions, which are counterparts of earlier known AdS pp-wave solutions with logarithmic modes, do not exist in the BI-NMG model.

Next, we have explored the field equations of the extended  NMG with  finite order curvature corrections, obtaining them from those of BI-NMG by performing a Taylor expansion and putting the result in  the form of a  massive Klein-Gordon type equation as well. Again, comparing the resulting equation with that obtained  for NMG, we  have found that the  general type N solution of NMG is  also subject to this model, with an appropriate adjusting of the parameters. We have also found that the model under consideration admits the counterparts of  the critical point solutions of NMG, which  are not allowed in BI-NMG.  We have given the analysis of these solutions for various {\it even} and {\it odd} numbers (including sufficiently large ones) of  curvature corrections. Meanwhile, the existence of such solutions raises  a new challenging question  as to {\it what happens to the extension of NMG  with $N$-order curvature corrections if $ N\rightarrow \infty $  and $|x| \geq 1$}?  The critical point solutions for sufficiently large values of $ N $ indicate that the limiting theory is certainly not BI-NMG and one needs a modified extension of NMG in the limit $ N\rightarrow \infty $  and $ |x| \geq 1$.  We will continue the exploration of this question in our future works.

{\it Note added.} While writing this work, the  paper  of \cite{mtb} appeared where the type N solution of BI-NMG  is also discussed by using the general solution of \cite{ah3}.

\end{document}